# LYSTO: The Lymphocyte Assessment Hackathon and Benchmark Dataset

Yiping Jiao, Jeroen van der Laak, Shadi Albarqouni, Zhang Li, Tao Tan,
Abhir Bhalerao, Jiabo Ma, Jiamei Sun, Johnathan Pocock, Josien P.W. Pluim,
Navid Alemi Koohbanani, Raja Muhammad Saad Bashir, Shan E Ahmed Raza,
Sibo Liu, Simon Graham, Suzanne Wetstein, Syed Ali Khurram, Thomas Watson,
Nasir Rajpoot, Mitko Veta, Francesco Ciompi

*Abstract*—We introduce LYSTO, the Lymphocyte Assessment Hackathon, which was held in conjunction with the MICCAI 2019 Conference in Shenzen (China). The competition required participants to automatically assess the number of lymphocytes, in particular T-cells, in histopathological images of colon, breast, and prostate cancer stained with CD3 and CD8 immunohistochemistry. Differently from other challenges setup in medical image analysis, LYSTO participants were solely given a few hours to address this problem. In this paper, we describe the goal and the multi-phase organization of the hackathon; we describe the proposed methods and the on-site results. Additionally, we present post-competition results where we show how the presented methods perform on an independent set of lung cancer slides, which was not part of the initial competition, as well as a comparison on lymphocyte assessment between presented methods and a panel of pathologists. We show that some of the participants were capable to achieve pathologist-level performance at lymphocyte assessment. After the hackathon, LYSTO was left as a lightweight plug-and-play benchmark dataset on grand-challenge website, together with an automatic evaluation platform. LYSTO has supported a number of research in lymphocyte assessment in oncology. LYSTO will be a long-lasting educational challenge for deep learning and digital pathology, it is available at https://lysto.grand-challenge.org/.

This work was supported in part by European Union's Horizon 2020 research and innovation programme under grant agreement no. 825292 (ExaMode project, http://www.examode.eu), and from the Alpe dHuZes / Dutch Cancer Society Fund, grant number KUN 2014-7032. (Corresponding author: Francesco Ciompi.)

Yiping J., Jeroen L., Francesco C. is with Department of Pathology, Radboud University Medical Center, Nijmegen, The Netherlands (e-mail: ping@nuist.edu.cn , francesco.ciompi@radboudumc.nl, Jeroen.vanderLaak@radboudumc.nl). Yiping J. is also with School of Artificial Intelligence, Nanjing University of Information Science Technology. Jeroen L. is also with Center for Medical Image Science and Visualization, Linköping University, Linköpi. Shadi A. is with Helmholtz AI, Helmholtz Zentrum München, 85764 Nuerherberg, Germany, and also with Faculty of Informatics, Technical University of Munich, 85748 Garching, Germany. Zhang L. is with College of Aerospace Science and Engineering, National University of Defense Technology, China, and also with Hunan Provincial Key Laboratory of Image Measurement and Vision Navigation, China. Tao T. is with Macao Polytechnic University, Macao, China. Abhir B., Johnathon P., Navid A. K., Raja M. S. B., Shan E. A. R, Simon G., Nasir R. is with Department of Computer Science, University of Warwick, Coventry, United Kingdom. Jiabo M. and Sibo L. is with Huazhong University of Science and Technology, China. Jiamei S. is with Information Systems Technology and Design Pillar, Singapore University of Technology and Design (SUTD), Singapore. Josien P.W. P., is with Medical Image Analysis Group, Department of Biomedical Engineering, Eindhoven University of Technology, Eindhoven, The Netherlands. Suzanne C. W. and Mitko V. is with Medical Image Analysis Group, Department of Biomedical Engineering, Eindhoven University of Technology, Eindhoven, The Netherlands. Syed A. K. is with School of Clinical Dentistry, University of Sheffield, Sheffield, United Kingdom. Thomas W. is with GlaxoSmithKline Plc (GSK).

## I. Introduction

Cancer and the host immune system have a complex, yet not fully understood, interplay. Over the years, clinicians and researchers in immuno-oncology have been investigating mechanisms involved in the tumor-immune microenvironment (TME), aiming at designing biomarkers that can capture a snapshot of such a scenario, and use those biomarkers to address one of the stringent questions in oncology: what to do next?

Over the years, the role of immune cells, and in particular the tumor-infiltrating lymphocytes (TILs), has increasingly been investigated[1]. Within the context of TILs in histopathology, two main research lines can be identified. The first line relies on the analysis of standard hematoxylin and eosin (H&E) stained histopathology slides and the quantification of a *TIL score*[2], estimated as the percentage of tumor-associated stroma region covered by lymphocytes and plasma cells. Several studies have shown that such a TIL score has prognostic and predictive value in breast cancer [3] as well as across a number of cancer types[4].

The second line relies on immunohistochemistry (IHC) to analyze T-cells, a subset of lymphocytes. Using IHC, specific types of cells can be identified in histopathology slides by targeting them via antigen-antibody interactions, and using a specific chromogen to distinguish them from other cells. In the context of lymphocyte assessment, the Immunoscore[5] was promoted to focus on T-cells that are positive to CD3 (all T-cells) and CD8 (cytotoxic T-cells) IHC markers, in particular at the tumor invasive front and in the tumor bulk.

Both the Immunoscore and TIL scoring approaches assess the *density* of immune cells as a biomarker, which therefore relies on *counting lymphocytes* within a certain tissue region. This is a type of task that suffers from implicit variability and tediousness when performed by humans, such as pathologists, suggesting the potential value of computer-algorithms for lymphocyte assessment. However, despite the potentially simple nature of this task on IHC slides using computer algorithms, it has been shown recently[6] that detecting lymphocytes in IHC goes beyond simply "counting dark-brown spots". Moreover, in daily practice, IHC slides contain challenging regions such as dense clusters, possibly background staining, and presence of artifacts such as ink (see some examples in Figure 1). Additionally, similar to the well-known stain variation in H&E[7], IHC also suffers from variation in tissue preparation, staining and scanning that is implicitly present across different pathology laboratories.

With the Lymphocyte Assessment Hackathon (LYSTO) as well as the benchmark dataset, we proposed and fostered the development of computer algorithms for the automated quantification of CD3 and CD8 positive cells in subregions (i.e., patches) of digitized histopathological images of different cancer types, namely breast, colon, and prostate cancer. Hosted in 2019, this paper looks back at the organization, sample acquisition, and performance of developed frameworks during the event. We also reported recent progresses based on post-event submissions to our online platform.

Compared to previous challenges in this field, the LYSTO hackathon has two main novel aspects. First, it formulated the problem of cell count in a *weakly supervised learning* fashion, where a single label per image is provided, indicating the number of positive cells present in the image, rather than providing manual annotations of individual cell location. Second, it challenged participants to develop a solution in a short amount of time, namely a few hours, which justifies the 'hackathon' epithet, as well as its name, partly inspired by the word '*listo*', a Spanish term for 'clever', as well as 'ready/finished'.

Different from regular challenges in medical image analysis, LYSTO did not enforce specific restrictions on models, training schemes, or resource usage. As a one-day event in the form of hackathon or proof of concept, LYSTO encourages participants to focus on the problem, and try out any strategy that could be helpful. Some of the methods developed in the event have exceeded the performance of medical specialists, suggesting the potential of deep learning methods for quantitative analysis in clinics. During the 3-year post-event interval, the weighted kappa of the best submission has reached 0.9331, compared to the record of the on-site event (0.9270). LYSTO will serve as a long-lasting educational dataset for machine learning and computational pathology.

## II. RELATED WORKS

### A. Lymphocyte assessment in H&E slides

In histopathology, cell quantification made by human experts via visual estimation is known to suffer from intra- and inter-observer variability[8]. For this reason, recent studies proposed the use of deep learning to analyze digital pathology slides stained with H&E. In recent work on breast cancer research[9], high-TIL regions are recognized, and a deep learning model is then developed to quantify the TIL proportion. [10] developed a method to describe the spatial arrangement of TILs, which was shown to be correlated with the tumor recurrence in early-stage non-small-cell lung cancer.

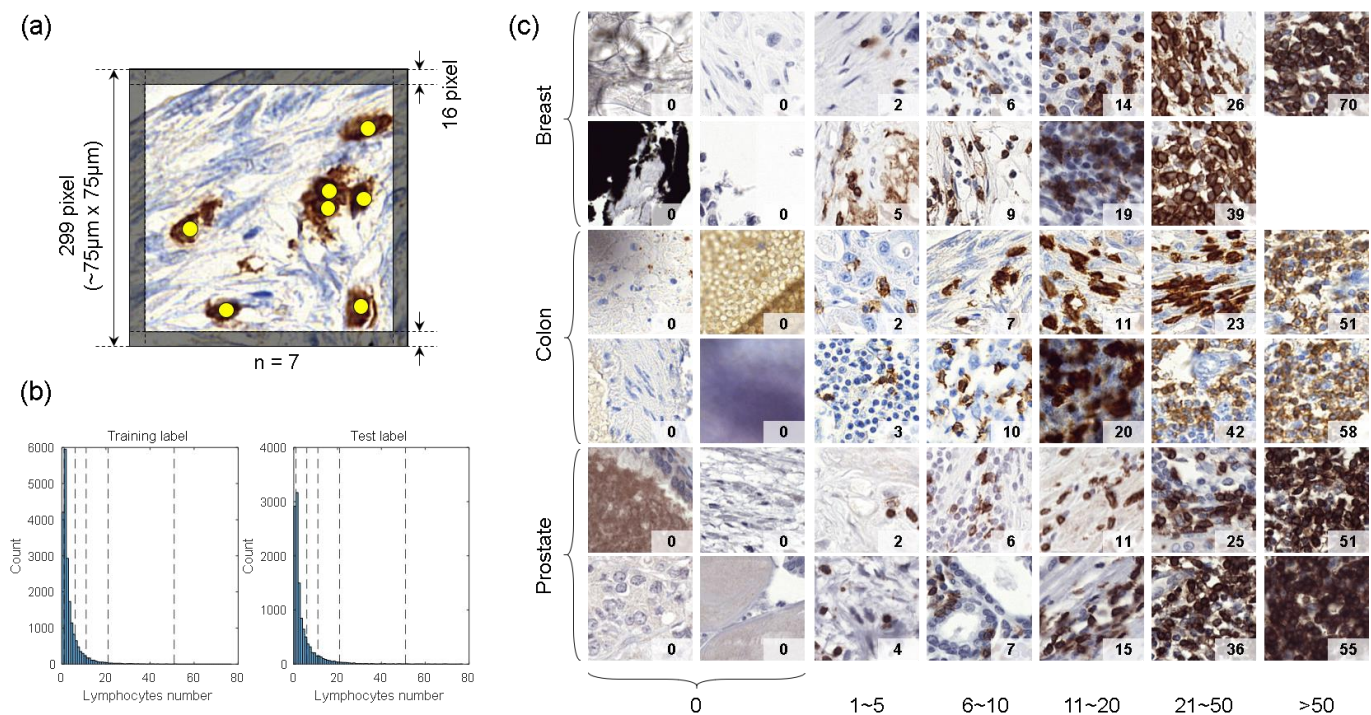

Fig. 1. LYSTO dataset. (a) Example image patch used in the experiment. The label is calculated as the positive cells number in the central 267x267 pixel. (b) Label distribution in training set and test set. (c) Sample examples used in the hackathon. The number at the right bottom of each image indicates the reference standard.

### B. Assessment of IHC slides

The composition of immune cells in a tumor is known to exhibit high heterogeneity [11], and sub-types of immune cells are infeasible to be identified solely in H&E slides. 3,3' diaminobenzidine (DAB) is widely used in IHC workflows, which makes the targeted antigens highlighted with brown color. Using specific targeted antigens, IHC is able to distinguish detailed sub-population of cells. Regarding the computer algorithms for cell detection and quantification in IHC, most works considered the quantification of the Ki-67 proliferation index. These methods can be categorized as color deconvolution-based methods and deep learning-based methods.

The color deconvolution transforms the conventional red-green-blue (RGB) images to the hematoxylin, eosin, and DAB channels in the optical density space[12]. Afterward, positive cells can be recognized from the DAB channel using specific thresholds[13]. Many variant methods have been proposed for better recognition accuracy, for example, using adaptive thresholds for better generalization [14], advanced classifier for brown/non-brown classification [15], utilizing super-pixel for cells of complex shape [16], and spatial correlation clustering for segmentation[17].

Similar to color deconvolution, some deep learning-based models obtain a segmentation map for positive nucleus, which can be used for cell counting [18], [19]. Other models are trained with single-nuclear images, aiming at a positive/negative classification task [20] or determining the strength of positive staining [21].

### C. Quantifying $CD3^+$, $CD8^+$ cells

CD3 and CD8 are markers for general T-cells and cytotoxic T-cells, respectively. Both CD3 and CD8 are membrane markers, leading to a brown ring-shaped membrane in positive cells [6] (see Figure 1). Several methods have been presented for detection and quantification of CD3 and CD8 positive cells in IHC slides. One approach is based on QuPath [22], which is an open-source software that can support IHC quantification. Similar to the method used for Ki-67 quantification, QuPath uses conventional color deconvolution to separate the Hematoxylin channel and DAB channel, followed by a peaks-finding algorithm to recognize single negative and positive cells.

Current deep learning-based methods for CD3-positive ($CD3^+$) and CD8-positive (CD8+) cell quantification can be categorized as pixel-level segmentation methods and cell localization methods[23]. Inspired by the ring-shaped morphology of the positive cells, a multi-class segmentation framework was proposed to label the center, nucleus, membrane, and background pixels in IHC slides[6]. The method was simplified to three labels (nucleus, membrane, and background) in a more recent study [23], achieving even better performances.

The cell localization models are trained to predict the spatial coordinates of positive cells directly, without the need to separate individual cells in segmentation maps. This type of methods include the locality sensitive method [24] and YOLLO [25]. [23] shows that a U-Net[26]-based model could lead to better accuracy than the aforementioned localization methods, especially in images with artifacts and cell clusters.

### D. The LYON Challenge

A challenge named LYON[23] was proposed for a similar problem as LYSTO, namely counting the positive lymphocytes in IHC images. LYON consists of 441 regions of interest (ROI), which can serve as a platform for existing algorithms; however, neither training nor test labels are available. The image source of LYSTO came from a subset of LYON. Differently, the samples in LYSTO are well-prepared, with a specific reference standard. To some extent, LYSTO can be seen as a twin challenge of LYON, with a concentration on a regression task for cell counting.

## III. LYSTO HACKATHON

In this section, we describe the data used in LYSTO, we present the design of the experiment and we describe the infrastructure made available to the participants, namely computing infrastructure and evaluation platform.

### A. LYSTO Datasets

We collected data from 83 digital pathology whole-slide images (WSIs) of colon ($n$=28), breast ($n$=33) and prostate ($n$=22) cancer slides. The image source of LYSTO came from a subset of the previous LYON challenge. The images were collected in a multi-centric fashion from $n$=9 medical centers in the Netherlands. For each case, slides with thickness of 2-4μm were cut from the tumor tissue block, stained with either CD3 or CD8 immunohistochemistry, and digitized using a Pannoramic 250 Flash II scanner (3DHistech, Hungary), resulting in WSIs with a pixel size of 0.24 μm/pixel.

For each WSI, approximately eleven regions of interest (ROIs) per slide were drawn and CD3 and CD8 positive cells were manually annotated exhaustively, as described in [23]. Particular care was taken to include ROIs from multiple regions of the WSI, containing a variety of lymphocyte patterns, including a) clusters of lymphocytes, b) groups of few or isolated T-cells, as well as c) artifacts such as background staining and regions with ink in the slide.

As done in the LYON paper, we split the set of slides into a training set ($n$=43 slides) and a test set ($n$=40 slides). The training slides were derived from two medical centers, and the test slides were derived from eight centers. Data from one lab was shared across the training and test sets, but no overlap of data from the same patient was present in the two sets.

### B. Sample Extraction

From each region of interest, we extracted partly-overlapping patches of size 299 × 299 pixel at full resolution (approximately 75×75 μm tissue region), using a stride of 200 pixels. This specific patch size was selected based on recent trends in computer vision [27], where a similar size is used as standard input size for well-known convolutional neural network architectures pre-trained on the ImageNet dataset. In the context of a hackathon, where time is a vital factor, starting with pre-trained models can reduce the time to model convergence. Therefore, providing an input patch size that is already compatible with most pre-trained models 'off-the-shelf' contributes to development of effective models in a short time period.

After collecting the patches, the label of a patch $y$ is defined as the number of annotated lymphocytes within it. To account for border effects, where a point annotation could be present very close to the border, but the associated cell only partly visible, we ignored annotations present within a border of 4 μm thickness (i.e., approximately half the average size of a T-cell) at the border of a patch (Figure 1). We defined several bins to discrete lymphocytes counts, namely 1~5, 6~10, 11~20, 21~50, 51~200, and >200. In the generation of the training set and test set, we balanced the labels according to these bins if possible. We also collected a number of patches without the presence of lymphocytes. To challenge participants with the dye artifacts in real-world applications, patches with $y=0$ were generated selectively according to the *brown score* proposed by [25].

TABLE I LABEL DISTRIBUTION OF LYSTO DATASET

| Properties | | Training | Test |
|---|---|---|---|
| No. of slides | Breast | 18 | 15 |
| | Colon | 13 | 15 |
| | Prostate | 12 | 10 |
| Label value | Min | 0 | 0 |
| | Max | 70 | 77 |
| | Mean | 3.11 | 3.92 |
| No. of sample | 0 | 4,208(21%) | 2,915(24%) |
| | 1~5 | 12,586(63%) | 6,663(56%) |
| | 6~10 | 2,008(10%) | 1,260(11%) |
| | 11~20 | 900(5%) | 790(7%) |
| | 21~50 | 290(1%) | 323(3%) |
| | 51~200 | 8(~0%) | 49(~0%) |
| | >200 | 0(0%) | 0(0%) |
| | Total | 20,000 | 12,000 |

Given the slides in training set and test, we randomly selected n=20,000 and n=12,000 patches, respectively according to the rules above. In addition to the patch and corresponding label, the cancer type are also recorded as optional information in the training set.

### C. External Validations

In parallel with data collection for the LYSTO hackathon, we also collected a set of $n$=10 lung cancer cases from Radboud University Medical Center. The training set and test set of LYSTO do not contain images from lung specimens. Therefore, this external validation set can be used to assess the robustness and the generalizability of the developed methods when applied to data from a different organ and scanned with a different scanner. All these slides were stained with a CD8 marker to detect cytotoxic T-cells and scanned with a Pannoramic 1000 scanner (3DHistech, Hungary), resulting in WSIs with a pixel size of 0.24 μm/pixel. Using the similar way as sample generation in LYSTO, we created $n$=54 ROIs created and gathered corresponding annotations.

Additionally, we considered the full ROIs used in the LYON test set. Despite being a super-set of the patches used in the test set of LYSTO, applying presented methods on larger ROIs from LYON allows to test the generalizability of methods' performance beyond the small context of single patches, when larger portions of challenging regions such as artifacts are present. Furthermore, running models on LYON data allows to compare models' performance with the human performance based on the results of the observer study conducted in [23], which involved four pathologists.

In order to perform validation on both external datasets, participants were asked by the hackathon organizers to run their methods on those external datasets within a few months after the experiment was completed. For this purpose, a Python script that can processing patches in larger ROIs with sliding window was provided.

### D. Timeline

The LYSTO experiment was a single-day event, held in conjunction with the Computational Pathology Workshop (COMPAY) at the MICCAI 2019 conference in Shenzhen (China). The hackathon was organized based on three main steps.

First, approximately one month before the day of the event, a small dataset of $n$=4,000 labeled patches was released publicly via the LYSTO website. The aim of this *pilot* dataset was to let potential participants get familiar with the type of data that will have been used during the event, the format used to release the dataset itself (e.g., HDF5 format), and start coding pipelines that could be reused and further developed during the event. The size of this dataset was defined by the organizers, being considered as large enough to be representative of variation in patch appearance yet not large enough to allow training an accurate model.

Second, the final official training set containing $n$=20,000 patches was released via the hackathon website three days before the day of the event. This time was chosen to allow participants to download the dataset before travelling to the location of the event.

Finally, the test set containing $n$=12,000 patches was solely released *on-site* via external storage units, which were manually distributed to participants. After the event, both training and test set were released publicly via the Zenodo platform. (https://zenodo.org/record/3513571)

### E. Rules

As an application-driven challenge, LYSTO encourages participants to try out various solutions.; therefore, no specific restrictions were enforced on model architecture, training schemes, data usage, or computing resources. We reformulate the problem of cell counting as a classification problem using pre-defined bins. This means that participants can solve the problem by using either classification, regression, or detection frameworks. Meanwhile, no restrictions were imposed regarding the data. Participants were allowed to reuse any materials in the community or append their in-house annotations. In summary, LYSTO is a open challenge, in which one can explore the most effective direction for future investigation towards cell counting task in IHC image.

### F. Performance Metrics

In order to make the metric compatible with classification, regression, and detection frameworks, we make LYSTO as a patch classification problem defined by counting. As stated above, all the patches were tagged with a group label, indicating whether 0, 1~5, 6~10, 11~20, 21~50, 51~200, and >200 positive cells are presented in the field of view. In practice, pathologists will not recognize and count every single cell in whole-slide images, and the intervals conforms to pathologists habits. To measure the consistency with reference standards, and penalize distinct prediction errors (e.g., predict a patch with 50+ positive cells as none), we use quadratic weighted kappa (QWK) coefficient as the main performance metric on the LYSTO test set and external lung validation set.

Moreover, since the intervals defined in LYSTO is same as that in LYON, we are able to compare our results with the reader study described in LYON[23]. For this purpose, we report sensitivity in the LYON test set.

### G. Baseline Results

To ensure the effectiveness of submitted methods, we provided a baseline prior to the event. The flowchart of the baseline is shown as Figure 2, which was built with following steps.

First, a color deconvolution algorithm[12] was applied to all the patches, producing three channels, namely the hematoxylin, eosin, and 3,3'-Diaminobenzidine (DAB). Second, we retrieved the intensity of the DAB channel and extracted the following statistics: maximum, minimum, and mean intensity value, standard deviation, and percentiles (2%, 5%, 10%, 20%, 50%, 80%, 90%, 95%, 98%) are calculated. Finally, a classification and regression tree (CART) model was trained to predict the lymphocytes number presented in each image, followed by model pruning. The CART model was developed using MATLAB (The MathWorks Inc., MA)

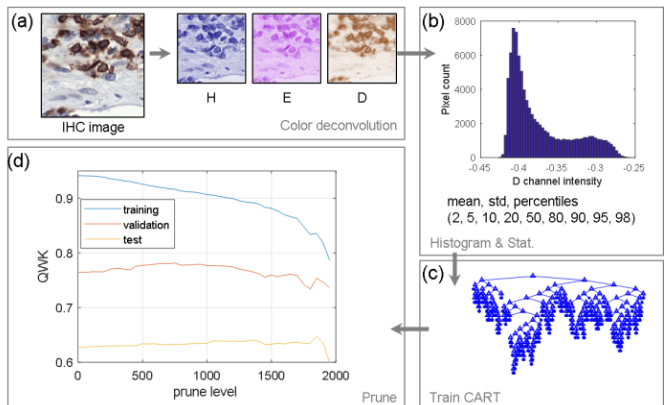

Fig. 2. Baseline development. (a) Color deconvolution; (b) Patch-level DAB channels statistics; (c) CART built with MATLAB; (d) Model performance and hyperparameter tuning.

The performance would range from about 0.628 to 0.649 when enumerating different prune levels. In the end, a prune level of 1800 was used for the released baseline, which got 0.635 on the LYSTO test set. This result was made available to the participants before the day of the hackathon, and the description of the procedure above was made available via website.

### H. Computing Resources

During the on-site event, we provided participants access to a dedicated GPU and storage on a cloud-based NVIDIA DGX-1 device, sponsored by NVIDIA, which supported the LYSTO hackathon. The official training set and test set were pre-loaded to the storage of the DGX-1 before the event. Additionally, participants were allowed to use local resources (e.g., their own laptop) as well as remote computing resources such as the server of their university without any restriction.

### I. Evaluation Platform

The hackathon organizers implemented an automatic evaluation procedure, which was made available on the grand-challenge.org web platform (https://lysto.grand-challenge.org/) to all the participants along with the released dataset. The participants were required to submit their predictions with a single file in CSV format, and the corresponding QWK score was calculated automatically. Examples of the submission format were also provided upfront.

## IV. METHODS

During the LYSTO hackathon, participants had the opportunity to create teams. In the end, five teams attended the on-site event and submitted their algorithm. After the event, we asked the team leaders to provide a short description of the developed algorithm. In this section, we provide a description of the main aspects of developed methods, namely 1) pre-processing, 2) data split, 3) model architecture and 4) training strategies.

### A. Team 1 (GSK)

*Preprocessing*: A 16-pixel border around each image was removed, resulting in an input size of 267×267 pixels. The RGB images were normalized to match the ImageNet statistics. During training, every training example was flipped horizontally and or vertically with probability $p = 0.5$ for each translation. Additionally, the contrast and brightness were adjusted uniformly ±20% with a probability $p = 0.75$.

*Data split*: The 20,000 training patches were stratified by the categories defined in the experiment, and they were further randomly split into a 75% training group and 25% validation group using the built-in scikit-learn method.

*Model architecture*: The model is a multi-task network that have a ResNet-50[28] backbone pre-trained on ImageNet. The last two layers of the ResNet-50 model were replaced with a CNN consisting of 4 convolutional layers, having 2048, 1024, 512, and 256 filters, respectively. These 4 layers all have a kernel size of 3, padding of 2, stride of 1, and dilation factor of 2. Afterward, task-specific branches for regression and classification tasks were added on the top of this CNN. The two branches are both consisted by cascaded layers including adaptive pooling layer, flatten layer, Batchnorm1D layer, dropout layer ($p=0.25$), ReLU activation, linear layer (64 neurons), Batchnorm1d layer, dropout layer ($p=0.25$), and the final linear layer (with 1 or 7 neurons, for regression and classification task, respectively).

*Training*: The Adam optimizer[29] was used to optimize the network using an adaptation of the one-cycle learning rate scheduler[30] implemented in fast.ai. The model was trained in two phases. In the first phase, the weights of the backbone layers were frozen and only the CNN and outputs task-specific layers were trained for 20 epochs using a maximum learning rate of 1e-3. In the second phase, the backbone layers were unfrozen and trained with the rest of the network for an additional 25 epochs. The maximum learning rate for the backbone is 1e-6. The learning rate for the CNN and output task-specific layers is lowered to 1e-4. Each epoch was scored on the validation set with the quadratic weighted kappa score obtained from the classification task. The best epoch from the first phase was used for the training in the second phase. The regression output from the best epoch of the second phase is used for the final prediction.

### B. Team 2 (HUST)

*Preprocessing*: Only the center 267×267 pixel region of the original 299×299 pixel image was used, because the provided labels do not count the edge pixels. The image data was augmented by random image flipping along the vertical or horizontal axis, rotating in $n \times 90°$, and perturbating the brightness within a small range of values.

*Data split*: The 20,000 training patches were divided into ten folds at random. The original intention was to pick the best model using ten-fold cross-validation, where nine folds serve as a training set and one as a validation set. However, during the hackathon it was decided to average the outputs of the 10 models trained during the cross-validation.

*Model architecture*: The model is a tailored ResNet-101[28] network for regression task. The last layer of the ResNet-101 network is removed, and a series of layers, including a global max pooling layer, a fully-connected layer (of 64 neurons) with ReLU activation, and a single-channel output layer, are attached.

*Training*: The output of the model is the predicted number of suspicious cells in the image. The mean squared error is implemented to measure the distance between predictions and given labels. The model was optimized with the Adam method [29] (learning rate = 0.001, step decay of 0.1 every 1500 iterations; exponential decay rates of 0.9 and 0.999 for the two moment) for a total of 6000 iterations with a batch size of 64.

### C. Team 3 (TIA Warwick)

*Preprocessing*: The images are firstly padded with reflection manner, resulting a spatial shape of 302×302. The pixel RGB intensity is normalized within the range [0,1]. During training, flipping, contrast, brightness, median blur, Gaussian blur, and Gaussian noise are used for data augmentation.

*Data split*: Our method contains two networks, trained with segmentation and regression tasks, respectively. For the segmentation pretraining, we fully annotated the pilot train dataset using the Automated Slide Analysis Platform (ASAP) software. In each image patch, we ensured that there is an agreement between the number of annotated lymphocytes and the reference standard count. This dataset was split by the ratio of 7:3 into training and validation sets. Part of the segmentation network was reused in the final regression network, which was trained with the on-site 20,000 training images using five-fold cross-validation.

*Model architecture and Training*: Initially, a HoVer-Net [31] model was trained to perform instance segmentation of positively stained lymphocytes. The model was trained in two stages. In the first stage, the ResNet-50 [28] encoder was initialized with weights pre-trained on ImageNet, and only the decoders were trained. In the second stage, both the encoder and decoder weights were updated. The segmentation model was trained using Adam optimizer with an initial learning rate of 1e-3 and a batch size of 8 on each GPU.

After training the HoVer-Net for lymphocyte segmentation, the decoders are removed, and a series of 3×3 convolution and max-pooling layers are added. Then, a global average pooling layer and a 1×1 convolution layer are added to output a single value that regress the number of positively stained lymphocytes. In other words, the HoVer-Net encoder is used as a pre-trained network to assist the task of lymphocyte counting. The regression network is trained using Adam optimizer with an initial learning rate of 1e-3 and a batch size of 8 on each GPU. Mean absolute error is used as the loss function. The final prediction is given by the five models generated by cross-validation.

### D. Team 4 (TU/e)

*Data split*: The data split was done at the WSI level, ensuring that the validation and test set contain at least 1 WSI from each of the three tissue types. Out of the 43 unique WSIs, 32 were used for training, 5 for validation, and 6 as a test set.

*Model architecture*: The best single model has a SE-ResNeXt-50 architecture [32]. Models with the following architectures were also trained: NASNet [33], Inception-ResNet-v2 [34], Xception [35] and SE-Net-154 [32].

*Training*: All the models were pre-trained on the ImageNet dataset, and optimized with stochastic gradient descent with momentum (learning rate 0.01, momentum 0.9, cosine annealing decay) for 50 epochs. The batch size may differ based on the GPU memory (for SE-ResNeXt-50, Inception-ResNet-v2 and Xception, the batch size is 18; for NASNetlarge and SE-Net-154, the batch size is 8). The input to the network are the original patches of size 299×299 pixels, except for NASNetlarge where the images are zero-padded to the size of 331×331 pixels. The images were augmented by random translation, rotation, scaling, shearing, flipping, and shifting of the color channels. The submitted prediction were obtained by taking the median of the predictions of all single models.

### E. Team 5 (mi2rl)

*Preprocessing*: Images with fewer DAB staining were processed by stain normalization, where two set of normalized images were used for training. We used the center crop with size 267×267 as input. Images were augmented with random rotation, horizontal and vertical flip. To deal with cases of lymphocyte count of over 200 which were not present in the training data, we performed an additional data augmentation: we took random patches with high concentrations of DAB and used them to generate a random image which we labeled as having lymphocyte count above 200. We did this for images with lymphocyte counts above 50, with 0.2 probability to be augmented.

*Data split*: 16305 image patches were used as the training set, and the rest was used as the validation set. The training set and validation set are independent on whole-slide level. Instead of labeling the images with the exact number of lymphocytes, we grouped the image patches with a similar number of lymphocytes into several bins, ensuring that each bin has more than 50 image samples. Image samples within the same bin were with the same label and the mean of the lymphocyte numbers per bin was used as the prediction.

with different labels are predicted by each team, and the percentage of them that is misclassified (b), and examples of

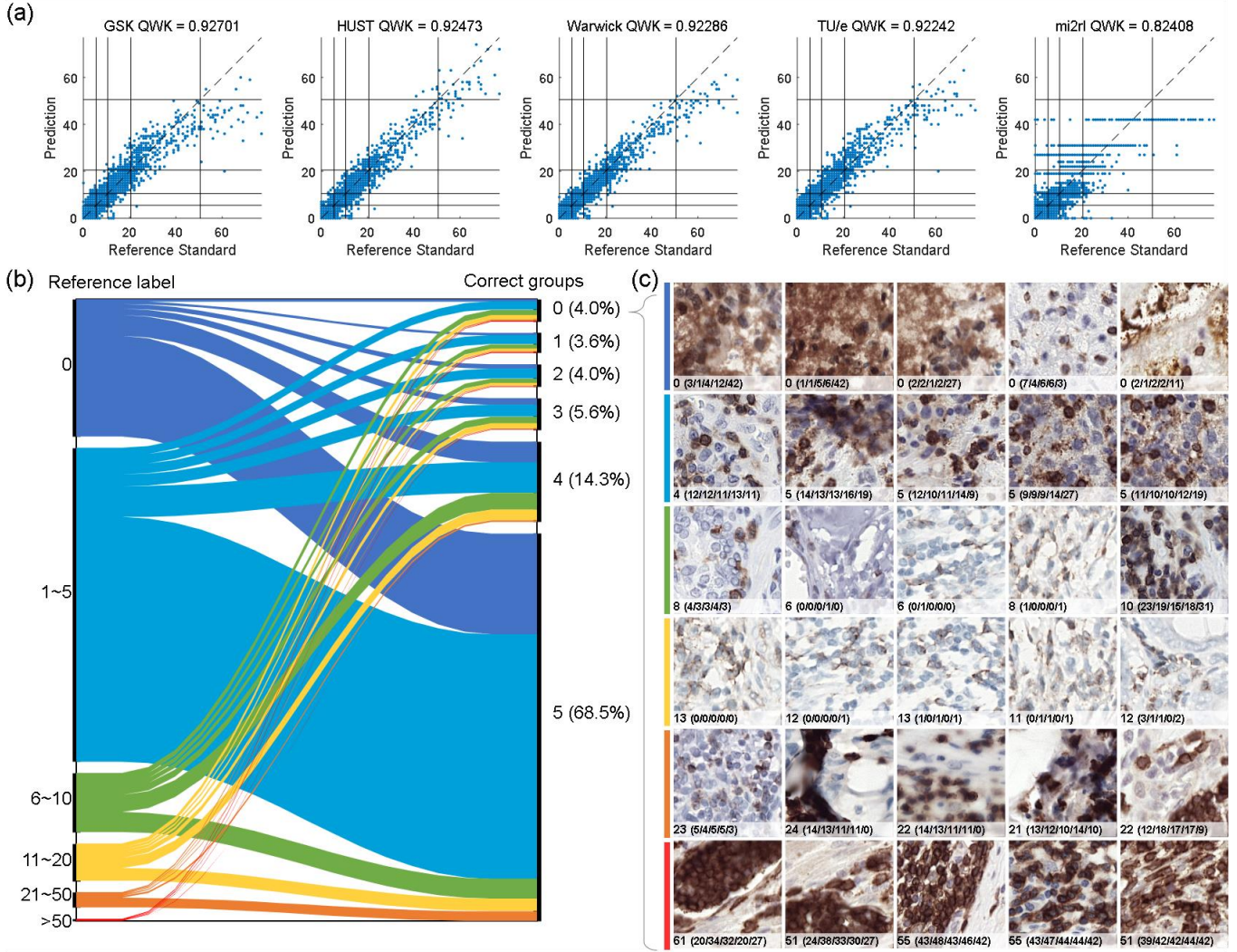

Fig. 3. On-site results. (a) Scatter plots of prediction versus reference standard. (b) Sankey diagram ground-truth versus number of groups that got correct prediction. (c) Examples of those 4% samples misclassified by all five groups.

*Model architecture and Training*: The model is a classification model with a DenseNet121 [36] backbone, and attached classification layers. For both two sets of stain normalized images, features generated from the last fully-connected layer of the backbone are fetched, concatenated, and fed into another fully-connected layer for classification. The model was trained using AdamW [37] optimizer (learning rate 1e-5, weight decay 0.05) for 10200 iterations with batch size 64. The loss was designed to minimize the distance between the median of the predicted and reference bins.

## V. RESULTS

### A. "On-site" results

The on-site results of the five developed methods are reported in Table I in terms of QWK scores. In Figure 3, we depict the scatter plots of counts predicted lymphocytes versus manual labels (a), the Sankey diagram showing how patches patches misclassified by all methods (c), grouped per category in each row, and indicating both manual and predicted labels

TABLE II LEADERBOARD OF LYSTO EVENT

| Team name | On-site | Rank | External (lung dataset) | Rank |
|---|---|---|---|---|
| GSK | 0.9270 | 1 | 0.9680 | 2 |
| TIA Warwick | 0.9229 | 3 | 0.9798 | 1 |
| HUST | 0.9247 | 2 | 0.8595 | 4 |
| TU/e | 0.9224 | 4 | 0.9652 | 3 |
| mi2rl | 0.8241 | 5 | 0.4678 | 5 |
| Baseline | 0.6350 | - | 0.8579 | - |

From the obtained QWK values and the scatter plots, we see that most methods achieved comparable performance, reaching a QWK>0.922, except for the mi2rl method, which obtained QWK=0.824. From the scatter plot, we can deduce that the main reasons for underperformance compared to the other methods are predictions of large number of lymphocytes in patches labeled as not containing lymphocytes, as well as a quantization effect on predicted labels, probably due to the label grouping strategy (i.e., binning) used by the team during training.

square error (MSE) loss or Huber loss.

### C. External Validation

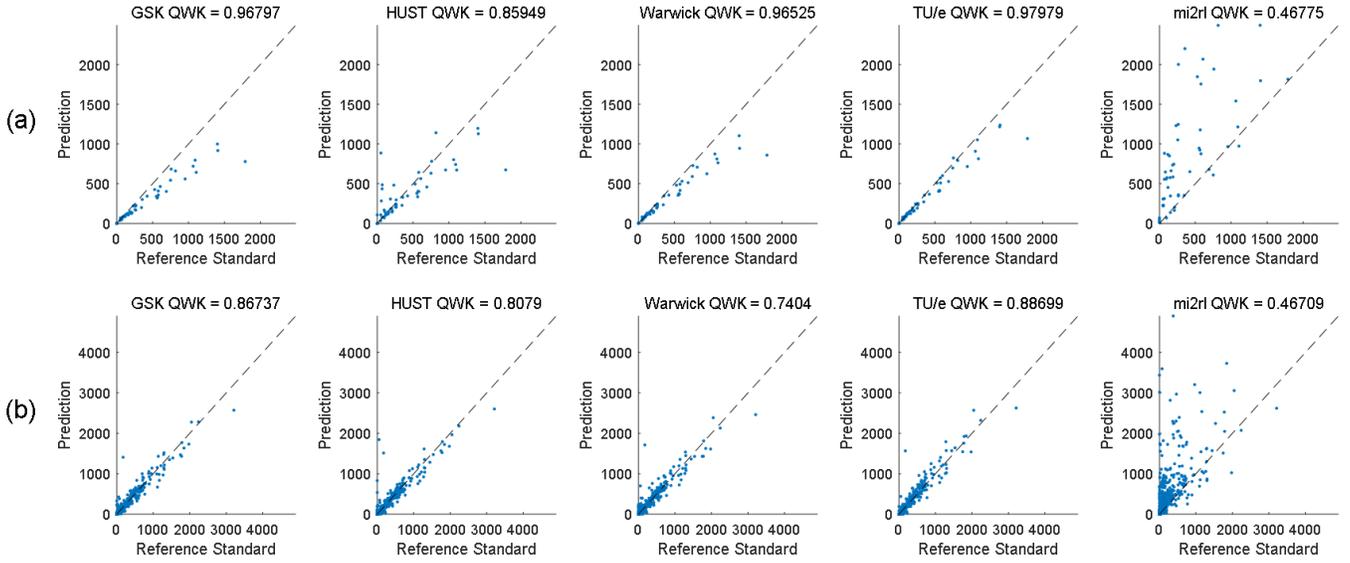

Fig. 4. Results of external validations. (a) The lung cancer cohort from Radboudumc. (b) The LYON test set.

According to the Sankey diagram, the samples with label 0 and 1~5 are relatively easy to be classified, as the majority of the two types of samples are correctly predicted by all the five teams. Such easy samples take 68.5% of the test set, however, 4.0% of the samples are misclassified by all the teams. Examples of these "difficult patches" are given in Figure 3, where we can see that misclassification is often correlated with the presence of background staining, resulting in a strong DAB signal but with the presence of little to no lymphocytes, which is still predicted by several methods as containing lymphocytes, in some cases up to 42 cells. Other examples include the presence of lymphocytes whose membrane is stained, but not completely, which were included during manual annotations because deemed to be positive to CD3 or CD8, but not picked up by most methods. Other reasons for misclassification are the presence of artifacts such as out-of-focus regions or ink, as well as the presence of clusters with a large number of lymphocytes, where most methods are capable to predict the presence of several lymphocytes, but overall tend to underestimate the exact number. This characteristic is also visible in the scatter plots of Figure 4, showing a trend of algorithms to underestimate the number of lymphocytes when it grows.

### B. Post-event Submissions

The LYSTO hackathon remains open for new submission after the MICCAI 2019 Conference. By now LYSTO has 667 registered users and receives 399 valid submissions. The highest QWK metric reaches up to 0.9331, which is higher than the 0.9270 achieved by the GSK group. According to the gathered method descriptions, these new submissions acquire similar techniques to the mentioned five on-site groups. Participants were likely to use ResNet-18, ResNet-50, ResNeXT or U-Net as their backbone. The task head was designed as classification, regression, or a combination of both. The classification head was normally trained by cross-entropy loss, whereas the regression head can be trained with mean

After the on-site event, the five submitted methods were evaluated on two additional datasets. In the two datasets, these methods are evaluated with sliding-window of 16 pixel overlap, and the count on ROI-level was reported.

*Lung dataset*: The first one was a set of *n*=54 regions of interest from *n*=10 whole-slide images of lung cancer resections stained with CD8. The results in terms of QWK are reported in Table II. Due to that the results are cumulated in larger field of view than a single patch, the region may contain more lymphocytes, but also more artifacts, compared to patches in the LYSTO datasets. In this sense, this part of experiment provides a broader overview of models' performance when applied to regions that are more representative of tissue morphology in whole-slide images used in routine clinical practice. Overall, the trend in performance is similar to what observed on the LYSTO test set, with most methods achieving QWK>0.9, and the mi2rl method underperforming, possibly due to the aforementioned effect of quantization. After ranking the methods based on performance, GSK remains the method with the best overall (in terms of averaged QWK of the on-site and the lung dataset) performance, followed by TIA Warwick, which achieved the best performance on the lung dataset.

TABLE III COMARISION WITH READER STUDY IN LYON

| Label | 0 | 1 ~5 | 6 ~10 | 11 ~20 | 21 ~50 | 51 ~200 | >200 | All |
|---|---|---|---|---|---|---|---|---|
| P1 | 0.78 | 0.11 | 0.25 | 0.15 | 0.32 | 0.71 | 0.54 | 0.41 |
| P2 | 0.96 | 0.17 | 0.20 | 0.15 | 0.27 | 0.58 | 0.73 | 0.44 |
| P3 | 0.78 | 0.28 | 0.15 | 0.20 | 0.32 | 0.48 | 0.35 | 0.37 |
| P4 | 0.96 | 0.33 | 0.25 | 0.15 | 0.55 | 0.65 | 0.43 | 0.47 |
| Average | 0.87 | 0.22 | 0.21 | 0.16 | 0.37 | 0.60 | 0.51 | 0.42 |
| [23] | 0.30 | 0.44 | 0.30 | 0.35 | 0.54 | 0.76 | 0.92 | 0.52 |
| GSK | 0.36 | 0.67 | 0.32 | 0.19 | 0.58 | 0.81 | 0.95 | 0.55 |
| TIA Warwick | 0.08 | 0.44 | 0.23 | 0.24 | 0.63 | 0.81 | 0.94 | 0.48 |
| HUST | 0.24 | 0.50 | 0.23 | 0.43 | 0.55 | 0.81 | 0.93 | 0.53 |
| TU/e | 0.04 | 0.28 | 0.23 | 0.14 | 0.52 | 0.78 | 0.95 | 0.42 |
| mi2rl | 0.16 | 0.06 | 0.00 | 0.05 | 0.22 | 0.42 | 1.00 | 0.27 |

*LYON Dataset*: Finally, we asked participants to run their methods on the full set of ROIs from the LYON test set (https://lyon19.grand-challenge.org/). These ROIs are publicly available, and were used to validate the lymphocyte detection methods presented in [23]. Notably, the test set of LYSTO is fully derived from the official test set of LYON, therefore, performance on full ROIs from LYON can be seen as a generalization of patch-level performance on LYSTO, one step closer to slide appearance observed in clinical practice (i.e., including more artifacts and cell clusters).

In this case, performance was measured in terms of sensitivity after assigning predictions to different groups. We report these result in Table III, where P1 to P4 stands for four pathologists involved in previous reader study [23]. The scatter plot of automatic count versus manual count for LYON ROIs in Figure 4. When applied to larger ROIs, we can observe that the method of TU/e reported higher sensitivity compared to the others, and most methods showed performance in line with the sensitivity of the best lymphocyte *detection* presented in [23], except for the intermediate category with 11-20 lymphocytes, where the detection method outperform both pathologists and LYSTO methods in terms of sensitivity.

Based on these results, we highlight two aspects. First, LYSTO methods trained in a weakly-supervised fashion to count lymphocytes by *estimating* their number achieve performance comparable to the ones of a lymphocyte detection method trained in a fully supervised fashion with manual annotations of lymphocyte locations. Second, in line with what observed with a detection method, LYSTO methods achieve performance that are in line or better than human performance at estimating the number of lymphocytes in regions of interest.

In particular, computer algorithms achieve higher sensitivity than pathologists when the number of lymphocytes grows, showing that limitations of the human visual system at exactly quantifying the number of cells when ROIs contain dozens or hundreds of cells. At the same time, humans can clearly distinguish positive cells from other regions with high-intensity of DAB signal, such as background staining or artifacts, which are challenging regions for computer algorithms, both for LYSTO and detection methods, as shown by the substantial difference in performance in the '0' category in Table III.

## VI. Discussion

According to the method descriptions, all five teams used deep learning methods that based on backbones including ResNet, SENet, and DenseNet. These solutions can achieve high QWK scores ranging from 0.8241 to 0.9270, which are highly consistent with human experts. Hosted in 2019, LYSTO shows that deep learning has already become the main trend in modern image analyses. Well-developed deep learning frameworks (e.g., Pytorch, Keras, etc.) allowed developers to set up a training and inference pipeline in a fairly short time frame. Meanwhile, an easy-to-use data format makes people focus on specific problems and spend less time on coding, interface, and debugging. LYSTO gives a typical example demonstrating that standardized data can lead to clinically applicable solutions in a very short time.

During the LYSTO experiment, no limitations were made about methodologies that could be used; therefore, we found that participants used a variety of machine learning strategies to improve their performance. For instance, GSK adopted a multi-task learning strategy, training the model with classification and regression tasks simultaneously. TIA WARWICK made extra manual annotations on a small part of the training set, and pre-trained a segmentation model using those annotations. Mi2rl adopted stain standardization method, and used fine-grained bins. TU/e used independently trained models with different architectures for ensembling, a technique that often leads to better predictive performance [38]. Both GSK and TIA WARWICK adopted a two-stage training strategy that has been widely used in transfer learning [39], [40]: in the first stage, a pretrained frozen backbone was attached to top layers that were trained; in the second stage, the entire network became trainable.

Based on the analysis of misclassified samples (Figure 3), we can observe that at least 4% of the samples can be considered as extremely hard to classify, since those were not correctly predicted by any of the groups correctly. These samples mostly contain background staining, artifacts, and brown debris in the field of view, which are often correlated with over-estimation of positive cells. At the same time, weakly stained cells, and those cells with partially stained membrane, may lead to under-estimation. Worth to be noted that, these cases are often present in histopathology practice, and can be addressed by human observers such as pathologists. However, computer models developed in this experiment and outside [23] are sometimes not capable to make a correct prediction and may tend to be guided mostly by the 'amount of brown' in the image.

Performance metrics in LYSTO are computed based on predictions of local fields of view (FoV), which are a tiny portion of what is assessed at whole-slide level in histopathology practice. The results on the external lung dataset and LYON dataset partly addressed this limitation, because they contain larger regions of interest, selected to be representative of a certain variety of tissue morphology at slide level. At the same time, they allow to further assess model robustness to larger regions coming from multiple centers (LYON) and a different organ (lung) than the ones used for training.

Based on the results for 54 ROIs in the lung dataset presented in Table I, most of the methods got a QWK larger than 0.85. Only mi2rl achieved relatively lower performance, which may be a result of accumulated error derived from their discrete predictions. Different from the lung dataset, the performance on LYON is defined by sensitivity, which is shown in Table III. The reason for this is the existence of performance measured on this set in a previous study [23], where a panel of pathologists was involved, which we adopted in this paper as well.

As shown, most automated counting methods have similar performance patterns regarding different counting intervals. Compared to human pathologists, these methods are prone to produce error in group '0'; however, when the number of objects is extremely large (>=200), the sensitivities can be higher than all of four involved pathologists. According to the averaged sensitivities, four out of five methods achieved better performance than the average of pathologists. The performance of TU/e and HUST is even comparable with fully-supervised method[6], [23].

It should be emphasized that, the participants of the LYSTO hackathon only had limited time to develop their methods for final submission; the rule is different from other challenges in digital pathology[23], [41], [42]. The results of the on-site LYSTO event indicate that one can develop a computer-aided diagnosis model that has pathologist-level performance. To some extent, the success can partially owe to the gathering of annotated digital slides, and a well-structured dataset. This preparation work made it possible to develop deep learning models in a 'plug-and-play' manner. Additionally, LYSTO introduced a problem of weak supervision, where only the number of objects is given, rather than their specific location. Participants have shown examples on how to address this problem via classification, regression, segmentation, or use other strategies like multi-task or relevant-task pretraining. Since our goal is to gather methods that can assist clinics, we did not enforce any restrictions to methodologies.

We made all LYSTO data publicly available, as well as the automatic evaluation platform on grand-challenge.org. The dataset is also available on Zenodo (https://zenodo.org/record/3513571). LYSTO has already supported a few recent research in oncology and medical image analysis[43], [44]. In this way, we envision LYSTO as a potential future benchmark for development in computational pathology, easy to access and process.

## VII. Conclusion

In this paper, we presented the summary of the Lymphocyte Assessment (LYSTO) Hackathon, which was held in conjunction with the 2019 Medical Image Computing and Computer Assisted Interventions (MICCAI) Conference. The aim of the hackathon was to develop automatic methods for immunohistochemistry quantification. We proposed the LYSTO dataset, which is composed of multi-center and multi-organ pathological images, as a reference to benchmark future computational pathology methods. Moreover, we left the LYSTO dataset as a long-lasting educational benchmark on https://lysto.grand-challenge.org/.

Acknowledgment

The authors would like to thank Nikki Wissink for her help in annotations of lymphocytes in lung tissue. Y. J. was funded by the China Scholarship Council for his internship in Radboud University Medical Center. Jeroen van der Laak is a member of the advisory boards of Philips, The Netherlands and ContextVision, Sweden, and received research funding from Philips, The Netherlands, ContextVision, Sweden, and Sectra, Sweden in the last five years. Jeroen van der Laak is chief scientific officer of Aiosyn BV, Netherlands. Francesco Ciompi is shareholder of Aiosyn BV, Netherlands, and received consultancy fees from TRIBVN Healthcare, France. Z. L. received fundings from the National Natural Science Funding of China (No.61801491) and Natural Science Funding of Hunan Province (No.2019JJ50728).

LYSTO has received funding from the European Union's Horizon 2020 research and innovation programme under grant agreement no. 825292 (ExaMode project, http://www.examode.eu), and from the Alpe dHuZes / Dutch Cancer Society Fund, grant number KUN 2014-7032.